\documentclass[12pt]{article}
\usepackage[dvips]{graphicx}
\def\jnfont{\rm}
\def\NPB#1,{{\jnfont Nucl.\ Phys.\ }{\bf B#1},}
\def\PLB#1,{{\jnfont Phys.\ Lett.\ B }{\bf #1},}
\def\PRD#1,{{\jnfont Phys.\ Rev.\ D }{\bf #1},}
\def\PRL#1,{{\jnfont Phys.\ Rev.\ Lett.\ }{\bf #1},}
\def\ZPC#1,{{\jnfont Z.~Phys.\ C }{\bf #1},}
\def\ETslash{\not{\hbox{\kern-4pt $E_T$}}}
\let\to=\rightarrow
\begin{document}
\thispagestyle{empty}

\hbox to \hsize{\hfil\vtop{\hbox{AMES-HET-00-05} 
		\hbox{July, 2000} }}
\vskip 15mm plus .5fil
\begin{center}
{\LARGE Searching for a stop-pair sample from top counting 
		 experiments at hadron colliders }\\
\vskip 10mm plus .5fil
 {\large Jin Min Yang $^a$, Bing-Lin Young $^{b,a}$ }\\
\vskip 7mm
$^a$  {\it Institute of Theoretical Physics, Academia Sinica,\\ 
	   Beijing 100080, China}\\
$^b$ {\it Department of Physics and Astronomy, Iowa State University,\\
	 Ames, Iowa 50011, USA}    

\end{center}
\vskip 30mm plus .5fil

\begin{center} Abstract \end{center}

The light stop if produced in hadron colliders in the form of 
$\tilde t_1 \tilde{\bar t_1}$ pair and
decaying through the likely decay chain  $\tilde t_1\to \tilde \chi^+ b$
followed by $\tilde \chi^+ \to \tilde \chi^0 f\bar f'$, can mimic closely
a top quark event when the mass of the stop is close to that of the top
quark.  Because of the much lower production rate, the stop event can be
buried under the top quark event sample.  In order to uncover the stop
event, specific selection cuts need to be applied.  Through Monte Carlo
simulation with suitable kinematic cuts, we found that such stop event can
be extracted from the top quark sample and detected by the top counting
experiments in the upcoming upgraded Tevatron and LHC.
However, because of the small statistics of the Run 1 of the
Tevatron, the stop signal remains hidden at Run 1.
\vfil
\newpage

\section{Introduction}
\label{sec1}

Search for SUSY particles is one of the primary tasks of the upgraded 
Fermilab Tevatron and the upcoming CERN Large Hadron Collider (LHC).
Because of the unknown masses of the sparticles and other free parameters,
various possibilities and strategies have to be considered in the search.
Among the plethora of sparticles, superpartners of the top quark, i.e.,
the stops, especially the lighter of the two mass eigenstates, denoted as 
$\tilde t_1$, is of particular interest. This is because that this stop 
has color interactions and is likely to be the lightest sfermion.  
Therefore, it could be produced in the gluon-rich environment of 
high energy hadron colliders.  The lightness of the stop is usually argued 
for the following reasons. Firstly, the large top quark Yukawa coupling can 
lead to a large negative one-loop contribution to the stop masses.  Hence, 
the stops could be significantly lighter than other sfermions at the 
electroweak scale due to the renormalization group evolution,
even if all sfermions have an universal mass at the unification scale.
Secondly, since the mixing between the sfermions corresponding to the 
left- and right-handed states of a given fermion is proportional
to the mass of the fermion, the large top quark mass can lead to a 
large mixing of the two stops.  This in turn causes a sizable mass 
splitting between the two mass eigenstates to make the lighter one, 
i.e., $\tilde t_1$, even lighter so as to be accessible to the current 
and future hadron colliders.  Thirdly, the existence of a light stop 
is preferred by electroweak baryogenesis~\cite{Carena}.  Finally,  on the
theoretical ground, the scenario that the first two-generation sfermions 
are as heavy as 10 TeV while the third generation sfermions are significantly 
lighter conforms with the naturalness principle~\cite{Dine}.

In the framework of minimal supersymmetric model (MSSM) with $R$-parity 
conservation, several possibilities of light stop searches from top
quark decay have been considered in specific scenarios in the literature.  
We recapitulate them briefly below. 

If the stop is the next-to-the-lightest super particle 
(NLSP), its only two-body decay mode is $\tilde t_1\to c \tilde \chi^0_1$
via loops~\cite{Hikasa}, where the lightest neutralino, $\tilde \chi^0_1$,
is assumed to be the lightest sparticle (LSP).
In the case that the stop is sufficiently light, one can consider
the exotic top decay $t\to \tilde t_1  \tilde \chi^0_1$ followed by
$\tilde t_1\to c \tilde \chi^0_1$.  Studies showed that this decay chain
in the $t\bar t$ pair events, if realized, can be observable in a large
part of the SUSY parameter space at the future runs of the Tevatron
collider~\cite{Hosch}.

Another possible decay mode of the stop in the case of $R$-parity
conservation, if the stop is light but heavier than the NLSP, is
$\tilde t_1\to b \tilde \chi^+_1$
through the tree-level coupling, where $\tilde{\chi}^+_1$ denotes the
lightest chargino. This decay mode will be the dominant decay channel 
of $\tilde t_1$ whenever it is allowed kinematically.
The phenomenology of the $t\bar t$ production followed by the decay chain
$t\to \tilde t_1 \tilde \chi^0_1$ and 
$\tilde t_1\to b \tilde \chi^+_1$ has been
studied soon after the observation of the top quark~\cite{Sender}.  
But now the significant higher values of the lower bounds of the masses of
$\tilde t_1$ and $\tilde \chi^0_1$, which
are given by about 122 GeV~\cite{CDF} and 45 GeV~\cite{LEP} 
respectively,  albeit under certain assumptions, make these top decay 
chains discussed above less likely. 

From the above discussion we see that the discovery of the light stop
through the top quark decay is not a promising possibility.  However, the
light stop offers a more direct route for its discovery at the Tevatron 
and LHC, i.e., the direct production of the stop pair, assuming the 
production cross section is sufficiently large and there exits a suitable 
decay channel for its identification.  We note that since the stop
pair production 
is a QCD process, the only uncertainty in the production cross section is
the mass of the stop and how it decays. 

If the stop is the NLSP and thus its only two-body decay mode is 
$\tilde t_1\to c \tilde \chi^0_1$, then the signal of the stop pair 
production at hadron colliders can only be two jets plus missing 
energy~\cite{Demina}. The large QCD background renders the signal 
impossible to uncover.
In this article, we examine, in the MSSM with $R$-parity conservation, 
the case of a stop with a mass close to that of the top quark and is 
heavier than the lightest chargino so that it decays dominantly through 
$\tilde t_1\to b \tilde \chi^+_1$, followed by
$\tilde \chi^+_1\to \tilde{\chi}^0_1 f \bar f'$ via a real or virtual
$W$-boson intermediate state, where $\tilde \chi_1^+$ is the lightest chargino
and $\tilde \chi^0_1$ the LSP.  Then a stop pair event will look like a top
quark pair event and can be easily masked by the latter~\cite{Berger}.
Through a detailed Monte Carlo simulation, the possibility of uncovering the 
possible stop pair events from the top quark counting experiment at the 
Tevatron 
and LHC colliders is investigated in detail.  As also being demonstrated
below, for a stop as light as the top quark, the stop sample will generally
be buried in the top sample at Run 1 because of the small statistics.  But
at the future runs of the upgraded Tevatron and at the LHC, such a stop
sample can be revealed through a series of suitable selection cuts.  

\section{Stop pair production and signatures}
\label{sec2}

Similar to the top quark pair production, in hadron collisions the stop
pair can be produced in the $q\bar q$ annihilation and gluon-gluon fusion
due to the $g\tilde t_1 \bar{\tilde t}_1$ coupling.
The lowest-order matrix elements will be used in our Monte Carlo simulation.
The absolute values squared of the two processes are given by
\begin{eqnarray}
\vert {\cal M} \vert^2(q\bar{q}\to\tilde t_1 \bar{\tilde t}_1 )
& = & 16 g_s^4 \frac{ \hat t_1 \hat u_1-m_{\tilde t_1}^2\hat s}{\hat s^2},\\
\vert {\cal M} \vert^2(gg\to\tilde t_1 \bar{\tilde t}_1 )
& = & 2 g_s^4\left[ 24 \left (1-2 \frac{\hat t_1\hat u_1}{\hat s^2}\right )
		    -\frac{8}{3}\right ] 
\left[1-2 \frac{m_{\tilde t_1}^2\hat s}{\hat t_1 \hat u_1}
\left(1-\frac{m_{\tilde t_1}^2\hat s}{\hat t_1\hat u_1}\right )\right ],
\end{eqnarray}
where $\hat s$ is the center-of-mass energy squared of the parton process,
$\hat t_1=\hat t-m_{\tilde t_1}^2$ and $\hat u_1=\hat u-m_{\tilde t_1}^2$ 
with $\hat t$ and $\hat u$ being Mandelstam variables. 
For parton distribution functions, we use CTEQ5L with 
$\mu=\sqrt {\hat s}$~\cite{CTEQ5L}. For a stop with a mass close to the top
quark, the QCD corrections enhance the total cross section of stop pair 
by a factor of $\sim 1.2$ at the Tevatron and $\sim 1.4$ at the LHC 
energies~\cite{zerwas}. 
These enhancements are taken into account in our calculation.

Although the QCD coupling in stop production processes is as strong as the
top quark production, the production rate of stop pair at a given energy
is much smaller than the top pair for similar masses.  The suppression of 
the stop pair production is caused largely by the fact that they are spin-0 
particles: 
(1) There is no sum over the spin projects of the final states that can
enhance the production rate by several fold.
(2) The $P$-wave coupling in the $q\bar{q}$ annihilation process give rise
to a $\beta^3$-dependence~\cite{zerwas} that caused the cross section to
be suppressed strongly near the threshold.
While the suppression factors work at all collider energies, 
the stop production at the Tevatron is suppressed more severely because the
dominant production of stop at the Tevatron is through the $q\bar{q}$
annihilation. 
  
As stated in the Introduction, we focus on the possibility of the
$\tilde{t}_1$ decay chain $\tilde t_1\to b \tilde \chi^+_1$ and
$\tilde \chi^+_1 \to \tilde \chi^0_1 f\bar f'$.  We assume a SUSY spectrum
in which all the sparticles involved in the decay chain are on-shell. We will
specify the relevant mass values of the sparticles below.  As pointed out
early, the two-body decay channel $\tilde t_1\to b \tilde \chi^+_1$ will
be dominant if the final state particles are on shell.  Thus in our calculation
we approximate the branching ratio of this mode as 100\%.  For the
subsequent 3-body decay of the chargino, we use its full matrix element
in our Monte Carlo simulation and take the total width of chargino to be
the sum of these 3-body decay channels of all allowed $f\bar{f}'$.  We
also assume that the charged Higgs boson, sleptons and squarks are much
heavier than the $W$-boson so that these three-body decays proceed 
dominantly through the $W$-boson intermediate state~\cite{Hikasa}.

So the stop pair  $\tilde t_1 \bar{\tilde t_1}$ production followed by the
decay chain $\tilde t_1\to b \tilde \chi^+_1\to b \tilde \chi^0 f\bar f'$
gives rise to 
top-like signatures except for two extra neutralinos which escape
detection.  There are three possible observing channels for the stop-pair
event: dilepton+2-jet, single lepton+4-jet and all(six)-jet.  All three
channels are associated with a significant amount of missing energies.  
The all-jet channel has the largest
rate but will subject to a very large QCD background, and thus is not
suitable for isolating the stop signal.  The dilepton channel has the
lowest rate and, furthermore, it is difficult to find a mechanism to
enhance the stop/top
rate to find the "smoking gun" of the stop pair production.  So we
will not use it, either. In the remaining single lepton+4-jet channel, the
best signal is $\ell+4j/b +\ETslash$ for the purpose of distinguishing the
stop event from a top pair event.  Here $4j/b$ represents a 4-jet event
with at least one of the jets passing the $b$-tagging criterion.  As is
shown below, we can find very effective selection cuts to enhance the
stop/top ratio for this signal.

\section{Relevant SUSY parameters}
\label{sec3}

There are several SUSY parameters involved in our calculation.
First of all, the stop mass is the most important parameter. We will fix 
it in the range of 170 GeV in most of our numerical examples.  But we will 
vary it
to find out how heavy it can be for the stop signal to be observable. For 
the neutralino and chargino masses, as well as their couplings, there are four
independent parameters: $M$, $M^{\prime}$, $\mu$ and $\tan\beta$.
$M$ is the $SU(2)$ gaugino mass and $M^{\prime}$ the hypercharge
$U(1)$ gaugino mass, $\mu$ the coefficient of the Higgs mixing term
in the superpotential, $\mu H_1 H_2$, and $\tan\beta=v_2/v_1$ the ratio of 
the vacuum expectation values of the two Higgs doublets. We work in the 
framework of the MSSM and assume the grand unification of the
gaugino masses, which gives the relation
$M^{\prime}=\frac{5}{3}M\tan^2\theta_W\simeq 0.5 M$.  This reduces the
independent parameters needed to three: $M$, $\mu$ and $\tan\beta$.  
For the three independent parameters, the chargino-neutralino sector can
be divided into two regions, the gaugino-like region ($M<|\mu|$)
and the higgsino-like region ($M>|\mu|$).

The gaugino-like region is favorable for the discovery of the stop signal.
In this region the lightest neutralino $\tilde \chi^0_1$ is mainly composed 
of the hypercharge $U(1)$ gaugino (bino), and the lightest chargino
$\tilde \chi^+_1$ is mainly composed of the charged  $SU(2)$ gaugino (wino).
So the $\tilde \chi^0_1$ mass is about half of that of the $\tilde \chi^+_1$. 
The large mass splitting between  $\tilde \chi^0_1$ and $\tilde \chi^+_1$
is needed to produce the required energetic jets or
lepton in the decay $\tilde \chi^+_1\to \tilde \chi^0_1 f \bar f'$, 
so that they can pass the necessary kinematic cuts.  

On the contrary, the higgsino-like region is unfavorable for the stop signal.
In this case, both the lightest neutralino $\tilde \chi^0_1$ and the
lightest chargino $\tilde \chi^+_1$ are mainly composed of the higgsino
fields. As a result, $\tilde \chi^0_1$ is almost degenerate with
(but lighter than)  $\tilde \chi^+_1$.   Then the lepton or jets
produced in the decay $\tilde \chi^+_1\to \tilde\chi^0_1 f \bar f'$ will be
too soft to pass our selection cuts.  So in this case the stop signal 
will be significantly reduced and likely hidden under the top
events even stop pairs are produced.

In our calculation we choose the following representative set of values
for the parameters in the gaugino-like region 
\begin{eqnarray} \label{para}
M=100 {\rm ~GeV}, \mu=-200 {\rm ~GeV}, \tan\beta=1. 
\end{eqnarray}
The chargino and neutralino masses in units of GeV are then given by
\begin{eqnarray}
& & m_{\tilde\chi^+_1}=120, ~m_{\tilde\chi^+_2}=220,~\nonumber\\
& & m_{\tilde\chi^0_1}=55,~m_{\tilde\chi^0_2}=122,~m_{\tilde\chi^0_3}=200,
~m_{\tilde\chi^0_4}=227.
\end{eqnarray}
As expected, $m_{\tilde\chi^0_1}$ is about half of $m_{\tilde\chi^+_1}$.

It should be remarked that SUSY parameters are generally not well-constrained
experimentally at the present time.  The only robust constraints are the
LEP and Tevatron lower bounds on some of the sparticle masses~\cite{report}.
In addition, the intermediate value of $\tan\beta$ is favored by 
low energy experiments~\cite{low}.  Therefore, the above SUSY parameter 
values used in our calculation are not the only choice.  They are a set of
representative values which are allowed by the current experimental
bounds and often applied for simulation.

\section{Selection of $\ell+4j/b +\ETslash$ events  }
\label{sec4}

In making the analyses, we simulate the energy resolution
of the detector by  assuming a Gaussian smearing of the energy of 
the final state particles, 
\begin{eqnarray}
\Delta E / E & = & 30 \% / \sqrt{E} \oplus 1 \% \rm{,~for~leptons~,} \\
	     & = & 80 \% / \sqrt{E} \oplus 5 \% \rm{,~for~hadrons~,}
\end{eqnarray}
where $E$ is in GeV, and $\oplus$ indicates that the energy-dependent and 
energy-independent terms are added in quadrature.  

The basic selection cuts are chosen as follows. 
For the Tevatron, the cuts are 
\begin{eqnarray}
p_T^{\ell}                   &\ge& 20 \rm{~GeV}~,\nonumber \\
p_T^{\rm miss}               &\ge& 20 \rm{~GeV} ~,\nonumber\\
p_T^{jet}                   &\ge& 15 \rm{~GeV}~,\nonumber\\
\eta_{jet},~\eta_{\ell}  &\le& 2.0 ~,\nonumber\\
\label{basic1}
\Delta R_{jj},~\Delta R_{j\ell} &\ge& 0.5 ~.
\end{eqnarray}
For the LHC, the cuts are chosen to be
\begin{eqnarray}
p_T^{\ell}                   &\ge& 20 \rm{~GeV}~,\nonumber\\
p_T^{\rm miss}               &\ge& 30 \rm{~GeV} ~,\nonumber\\
p_T^{jet}                    &\ge& 20 \rm{~GeV}~,\nonumber\\
\eta_{jet},~\eta_{\ell}      &\le& 3.0 ~,\nonumber\\
\label{basic2}
\Delta R_{jj},~\Delta R_{j\ell} &\ge& 0.4 ~.
\end{eqnarray}
Here $p_{T}$ denotes the transverse momentum, $\eta$ is the pseudo-rapidity,
and $\Delta R$ is the separation in the azimuthal angle-pseudo rapidity 
plane $(~\Delta R= \sqrt{(\Delta \phi)^2 + (\Delta \eta)^2}~ )$ between 
a jet and a lepton or between two jets.

For the signal, we require to tag at least one $b$ in $\ell+4j/b +\ETslash$.
The tagging efficiency is 53\% at Run 1 and expected to reach
85\% at Run 2 and Run 3~\cite{tev2000}. For the LHC we assume the tagging
efficiency to be the same as the Tevatron Run 2.  Under the above basic
selection cuts and $b$-tagging, the ratio of the top events
$\ell+4j/b +\ETslash$ to the QCD backgrounds is about 12:1~\cite{tev2000},
which we will use to evaluate the QCD backgrounds.

We noticed that for the top events and $W$+jets background events the missing 
energy comes only from the neutrino of the W decay, while for the stop events
the missing energy contains two extra neutralinos.  From the transverse
momentum of the lepton, $\vec P_T^{\ell}$, and the missing
transverse momentum, $\vec P_T^{\rm miss}$, we construct the transverse mass 
\begin{equation}
m_T(\ell,p_T^{\rm miss}) = \sqrt{ (|\vec P_T^{\ell}|+|\vec P_T^{\rm miss}|)^2
- (\vec P_T^{\ell}+\vec P_T^{\rm miss})^2}.
\end{equation}
As is well-known, if $P_T^\ell$ and $p_T^{\rm miss}$ are from the decay 
products of a parent particle, the transverse mass is bound by the mass 
of the parent particle.  For the top quark and $W$+jets
background events, where the only missing energy is from the neutrino of
the W decay, $m_T(\ell,p_T^{\rm miss})$ is always less than $M_W$ and peaks
just below $M_W$, although kinematic smearings can push the bound and the
peak above $M_W$.  For the stop events, there is no such peak due to
the extra missing energies of the neutralinos. The transverse mass
distributions of the stop and top quark events are shown in Fig.1.
The transverse mass distribution of the top quark events indeed conforms 
with the expectation, i.e., it peaks just below 80 GeV and significant
distribution appears above 80 GeV due to the smearing.
In order to substantially enhance the ratio stop/top, Fig. 1 suggests that
we apply the following cut,
\begin{equation}
m_{T}(\ell,p_T^{\rm miss}) \not\in 50\sim 100 {\rm GeV}.
\end{equation}

To further enhance the stop/top ratio, we construct four different
invariant masses, denoted as $M(3j)$ by using three jets out of the four
jets in the event.  We define the one which is closest to 175 GeV as the 
reconstructed top quark and denote its value as $M_{\rm top}(3j)$.  
The $M_{\rm top}(3j)$
distribution at the upgraded Tevatron energy is shown in Fig.2.
As expected, for the top quark events, there is a peak at the top quark
mass, $M_t=175$ GeV.  To enhance the stop/top ratio further, we suppress
the top quark events by selecting $M_{\rm top}(3j)$ to be 20 GeV away from the
top quark mass (175 GeV), i.e., 
\begin{eqnarray}
\vert M_{\rm top}(3j) - M_t\vert &\ge& 20 \rm{~GeV}.
\end{eqnarray}

\section{Numerical results} 
\label{sec5}

For the parameter values specified in Sec. \ref{sec3}, the
$\ell+4j/b +\ETslash$  
cross section from the stop and top quark events at the Tevatron 
and LHC,  under various cuts, are presented in Table 1.
The basic selection cuts Eqs. (\ref{basic1}) and (\ref{basic2}), 
which are necessary to reduce the QCD background, affect the stop cross
section more than that of the top quark and therefore lower the stop/top
ratios. This is because the lepton and jets from the stop events are
relatively softer and thus harder to pass the selection cuts, resulting
in the stop/top ratios below 5\%.  This makes the discovery of a stop
impossible if the systematic uncertainty is considered.  However, the
cut of the reconstructed top quark mass suppresses the top quark and stop
cross sections respectively by factors of about 12 and 1.6, drastically 
enhancing the stop/top ratio.  The transverse mass cut can further
increase the stop/top ratio by suppressing the top cross section by a
factor of about 3.3 and that of the stop by 1.9.

In extracting the new physics signal from the top quark events, 
various uncertainties have to be taken into account besides the experimental 
statistical and systematic errors.  The present uncertainty of the 
standard model $t\bar t$ cross section is at the 5\%\ 
level~\cite{Bonciani}.  Additional uncertainties that come from the error 
in $m_t$ will be much reduced with the expected more precise determination 
of $m_t$ (within 2.8 and 0.8 GeV as quoted for Run~2 and 3~\cite{tev2000}).  
For illustration we use the total systematic error of 5\%. Combined with 
the statistical error for each run, we obtain the total errors for the top
quark events as listed in Table 2. 
In estimating the statistical errors, the QCD backgrounds which are mainly 
W+jets have also been taken into account. Under the basic selection cuts 
they are reduced to about 1/12 of the top events in the channel of 
$\ell+4j/b+\ETslash$~\cite{tev2000}. The cut on the reconstructed top mass 
$M_{\rm top}(3j)$ cannot suppress the QCD backgrounds significantly and, 
therefore, to be conservative, we neglect such suppressions. However, the 
transverse mass cut is expected to suppress the QCD W+jets backgrounds 
significantly because the missing energy is only from the neutrino of the 
$W$ decay just like in a top event. So we assumed the transverse mass cut 
suppress the QCD backgrounds by a factor of about 3 as in the top quark case.

As indicated in Tables 1 and 2, although the stop/top ratio is enhanced 
significantly by the suitable cuts, Run 1 of the Tevatron is unable to       
observe the stop events because of the small statistics. So even for
the favorable case (gaugino-like region) under consideration, the stop
pair events will still be hidden in the top pair samples. 

For Run 2A with a luminosity of 2 fb$^{-1}$, the statistical error
is still large. As showed in Table 2, after combined with the 5\%
systematic error, the total error is 6\%, 14\% and 24\% under three
different selection cuts. Comparing with the stop/top ratio in 
Table 1, which is 4\%, 30\% and 49\% under the corresponding cuts,
we see that the $M_{\rm top}(3j)$ and $m_T(\ell,p_T^{\rm miss})$ cuts
drive the sensitivity to the $2\sigma$ level. This is still below 
the discovery limit which is usually required to be a $5\sigma$ deviation
or more.

For Run 2B (15 fb$^{-1}$) and  Run 3 (30 fb$^{-1}$), the statistical errors  
are significantly reduced, as shown in Table 2. For example,
comparing the total errors (5\%, 6\%, 8\% under the three selection cuts) at
Run 3 with the stop contributions (4\%, 30\%, 49\% under the three 
corresponding cuts), we conclude that the stop event under the 
$M_{\rm top}(3j)$ and $m_T(\ell,p_T^{\rm miss})$ cuts is observable 
($\ge 5\sigma$).  

For LHC, because of the large production rates, even for the
low luminosity run (say $10$ fb$^{-1}$), the statistical error is reduced
to be negligible.  So the total error under each selection cut is
dominated by the systematical error which is assumed to be 5\%.  Comparing
with the stop contributions (5\%, 40\%, 62\% under the three selection
cuts ), one sees that the stop sample after the $M_{\rm top}(3j)$
and $m_T(\ell,p_T^{\rm miss})$ cuts will undoubtedly be observable.  

In the above results of stop events we fixed stop mass to be 170 GeV.  
In Figs.3 and 4 we present the stop/top ratio versus stop mass under
the basic plus $m_T(\ell,p_T^{\rm miss})$ plus $M_{\rm top}(3j)$ cuts.
The horizontal dotted lines are the limits required by the discovery 
($5\sigma$), evidence ($3\sigma$) and (if not observed) exclusion 
($2\sigma$) of the production of stop pairs. We see that the LHC
(10 fb$^{-1}$) is able to discover a $135 \sim  215$ GeV stop, 
while Run 2B (15 fb$^{-1}$) of the Tevatron
can discover a $135 \sim 175$ GeV stop. If not discovered, a stop
lighter than $245$ ($200$) GeV will be excluded by LHC  (Run 2B of
the Tevatron) at $95\%$ C.L.. Of course, such results are valid only
for the gaugino-like scenario with the specific parameter values
we considered as given in Sec. \ref{sec3}

The peaks in Figs. 3 and 4 are the artifact of the cuts applied and
can be understood as follows. As the stop mass decreases, the stop pair 
production rate increases. However, the $b$-jet from 
$\tilde t_1\to \tilde \chi^+_1 b$ becomes softer and thus
harder to pass the selection cuts so as to decrease the ratio stop/top. 
At low values of stop mass, this latter effect is stronger and thus the 
net effect is to decrease the ratio stop/top  for decreasing stop mass.
At the high end of the stop mass, the phase space suppression of the 
production rate leads to decreasing stop/top for increasing stop mass. 
The balance of these opposite effects give rise to the peaks.
The peak will shift to higher value for higher parton center of
mass energy as shown in Figs. 3 and 4.     
   
\section{Summary and discussion} 
\label{sec6}

We have investigated the potential of the detection of the top-like events of
light stop pair at hadron colliders for the case that the mass of the
lighter stop is close to that of the top quark, and the SUSY parameters
lie in a range which allows the extraction of the stop signal.
Because of the much lower production rate relative to the top quark pair
production, the extraction of the stop event from the top sample requires 
special consideration to enhance the ratio stop/top.
Through Monte Carlo simulation with
suitable cuts, we found that a stop signal in the channel
$\ell+4j/b +\ETslash$ may be detectable in the top counting experiment 
in the upgraded Tevatron and LHC.  However, because of 
the small statistics, Run 1 of the Tevatron is unable to detect such a stop
presence, leaving the stop event hidden in the top pair sample
even if it is produced. 

We note that our calculation represents the results of a limited set of
numerical examples rather than the scanning of the whole SUSY parameter
space allowed.  Our results are dependent on the mass values of the
sparticles involved: the stop $\tilde{t}_1$, chargino $\tilde{\chi}^+_1$
and the neutralino $\tilde{\chi}^0_1$. To validate our analysis, the stop
mass must be larger than those of the lightest chargino and neutralino,
and their mass spectrum has to be gaugino like. Finally, 
as already stated in Sec. \ref{sec3}, we also note that if the 
lightest neutralino and chargino
are higgsino-like and thus their masses are close, the leptons and jets
in the final states would be too soft to pass our proposed isolation cuts.
Then, the stop signal would not be observable.  They will remain hidden
in the top pair sample even if the stops are produced at the upgraded
Tevatron and LHC.

It should be pointed out that throughout our analysis we worked in
the MSSM with $R$-parity conservation.  If R-parity is violated, there are 
also some interesting phenomenologies in the top-stop sector at
the Tevatron and LHC energies, some of which have been explored 
elsewhere~\cite{RV}. 

\section*{Acknowledgment}

JMY thanks Y. Sumino and C.-H. Chang for helpful discussions. 
BLY acknowledges the hospitality extended to him by Professor Zhongyuan Zhu
and colleagues at the Institute of Theoretical Physics, Academia Sinica, 
where part of the work was performed. 
This work is supported in part by a grant of Chinese Academy of Science
for Outstanding Young Scholars.

\newpage
\null\vspace{0.4cm}
\noindent
{\small Table 1:  Cross sections for $\ell+4j/b +\ETslash$, from stop and
top quark pairs at the Tevatron and LHC.  The basic cuts are given in
Eqs.(\ref{basic1}) and (\ref{basic2}).
The $M_{\rm top}(3j)$ and $m_T(\ell,p_T^{\rm miss})$ cuts are given by
$|M_{\rm top}(3j)-M_t|>20$ GeV and
$m_T(\ell,p_T^{\rm miss})\not\in 50 - 100$ GeV.
The stop events were calculated by assuming $M_{\tilde t_1}=170$ GeV,
$M=100$ GeV, $\mu=-200$ GeV and $\tan\beta=1$.
Tagging at least one $b$-jet is assumed for 53\% efficiency for the Tevatron
(1.8 TeV), $85 \%$ efficiency for the upgraded Tevatron (2 TeV) and LHC.
The 'No cut' column gives the result under the condition of b-tagging
in the absence of any kinematic cuts. 
The charge conjugate channels are included. }
\vspace{0.1in}
\begin{center}
\begin{tabular}{|c|l|c|c|c|c|} \hline
     & & & & &\\ 
     & & no cut& basic cuts & basic cuts               & basic cuts \\
     & &       &            & $+$                      & $+$ \\ 
     & &       &            &$M_{\rm top}(3j)$ cut         &$M_{\rm top}(3j)$ cut \\  
     & &       &            &                          & $+$ \\ 
     & &       &            &                          & $m_T(\ell,p_T^{\rm miss})$ cut \\ 
     & & & & &\\ \hline  
     & & & & &\\
   & stop (fb)       & 59  &$ 12  $&$ 7.4  $&$3.9   $ \\ \cline{2-6}
 Tevatron &  top  (fb)       & 750 &$ 317  $&$26.0 $&$ 8.0 $  \\  \cline{2-6}
 (1.8 TeV) &  stop/top (\%)   & 7.9 &$ 3.8  $&$28.5 $&$ 48.8 $  \\ 
     & & & & &\\ \hline
     & & & & &\\
   & stop (fb)       & 136 &$ 27 $&$ 17  $&$ 8.8  $ \\ \cline{2-6}
 Tevatron  &  top  (fb)       & 1652&$ 690 $&$ 57 $&$ 18 $  \\  \cline{2-6}
 (2 TeV)   &  stop/top (\%)   & 8.2 &$ 3.9  $&$ 30.0 $&$ 48.9 $  \\ 
     & & & & &\\  \hline
     & & & & &\\
   & stop (pb)       & 26  &$ 2.92 $&$ 1.94  $&$ 0.90  $ \\ \cline{2-6}
 LHC     &  top  (pb)       & 170 &$ 60.8  $&$ 4.86 $&$ 1.46 $  \\  \cline{2-6}
(14 TeV) &  stop/top (\%)   & 15.3&$ 4.8   $&$ 40.0 $&$ 62.0 $  \\   
 & & & & &\\  \hline  
\end{tabular}
\end{center}
\vspace{1cm}
\eject

\null\vspace{0.4cm}
\noindent
{\small Table 2:  Numbers of expected top quark events in the channel  
$\ell+4j/b+\ETslash$, together with the associated errors.
The estimated total error is obtained by combing the statistical errors 
and a 5\% systematic uncertainty. }
\vspace{7mm}
\begin{center}
\begin{tabular}{|l|l|c|c|c|} \hline
     & & & & \\ 
     & & basic cuts & basic cuts          & basic cuts \\
     & &            & $+$                 & $+$ \\
     & &            & $M_{\rm top}(3j)$ cut   & $M_{\rm top}(3j)$ cut  \\  
     & &             &                     & $+$ \\
     & &            &                     & $m_T(\ell,p_T^{\rm miss})$ cut \\ 
     & & & & \\	\hline 
     & & & & \\  
   & top events          &$32   $&$3   $&$1   $ \\ \cline{2-5}
Run 1 ($0.1$ fb$^{-1}$) 
& Stat. error (\%)&$18.5 $&$81.6 $&$141.4   $  \\  \cline{2-5}
& Total  error (\%)      &$19.2 $&$81.8 $&$141.5   $  \\  
     & & & & \\	\hline
     & & & & \\ 
   & top events          &$ 1380 $&$ 114  $&$ 36  $ \\ \cline{2-5}
Run 2A ($2$ fb$^{-1}$) 
& Stat.  error (\%)&$ 2.8  $&$ 13.3 $&$ 23.6 $  \\  \cline{2-5}
& Total  error (\%)      &$ 5.7  $&$ 14.2 $&$ 24.1 $  \\  
     & & & & \\	 \hline
     & & & & \\ 
   & top events          &$10350 $&$ 855  $&$ 270  $ \\ \cline{2-5}
Run 2B ($15$ fb$^{-1}$) 
& Stat.  error (\%)      &$ 1.0  $&$4.8  $&$8.6  $  \\  \cline{2-5}
& Total  error (\%)      &$ 5.1  $&$7.0  $&$10.0  $  \\  
     & & & & \\	 \hline
     & & & & \\ 
& top events          &$20700    $&$ 1710  $&$ 540 $ \\ \cline{2-5}
Run 3 ($30$ fb$^{-1}$)
& Stat.  error (\%)   &$ 0.7     $&$ 3.4   $&$ 6.1 $ \\ \cline{2-5}
& Total  error (\%)   &$ 5.1     $&$ 6.1   $&$ 7.9 $ \\ 
     & & & & \\	 \hline
     & & & & \\
& top events &$6.1\times 10^5$&$4.9\times 10^4$& $1.5\times 10^4$\\ \cline{2-5}
LHC ($10$ fb$^{-1}$)
& Stat.  error (\%)
	     &$0.1$          &$0.6$            &$1.2 $  \\  \cline{2-5}
& Total  error (\%)      
	     &$5.0$          &$5.0$            &$5.1 $ \\ 
     & & & & \\ \hline  
\end{tabular}
\end{center}
\vspace{1cm}

\begin{figure}[htb]
\vspace*{-2cm}
 \includegraphics[height=15cm,width=12cm, angle=90]{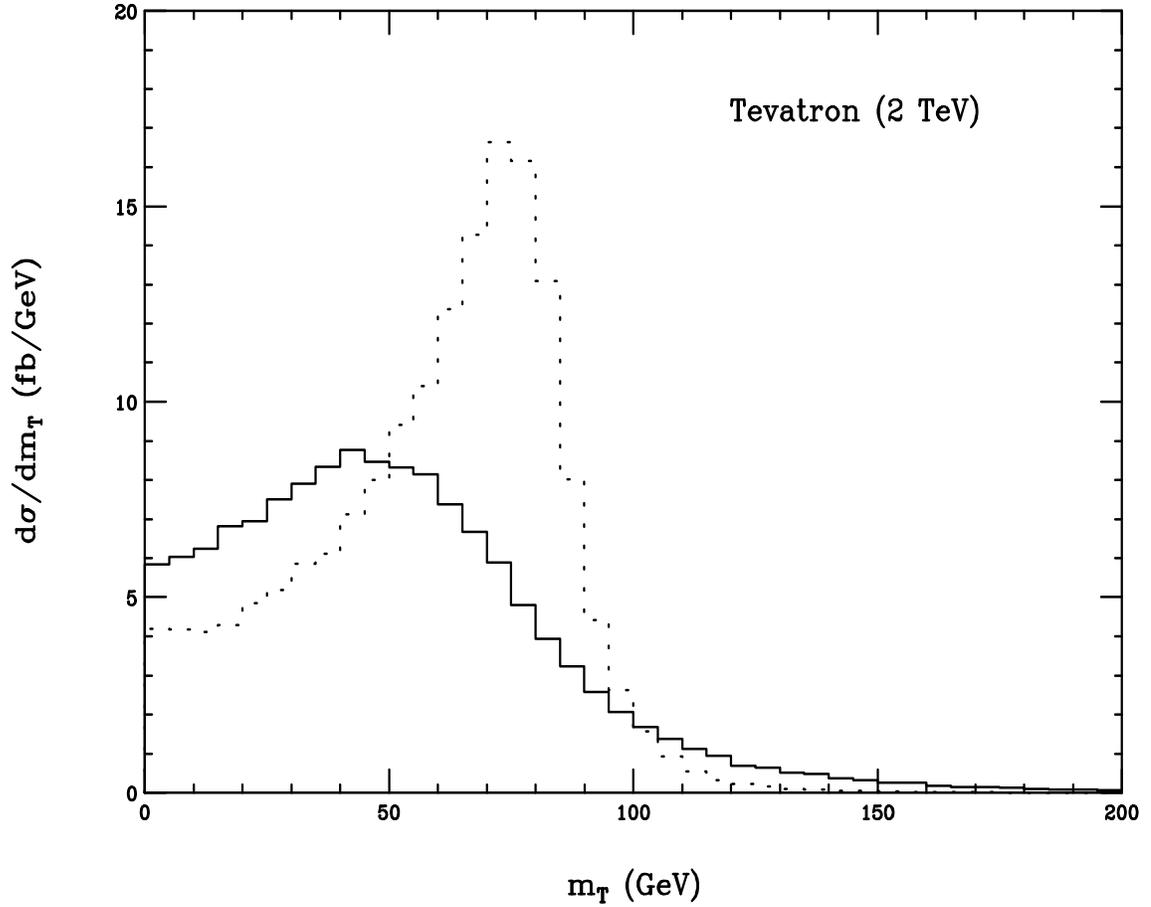}
\caption[]{ The transverse mass, $m_T(\ell,p_T^{\rm miss})$,
distribution of $\ell+4j/b+\ETslash$ at the Tevatron collider.
The solid curve is for the stop event with stop mass of $170$ GeV.  
The dotted curve is for the top quark event scaled down by a factor of 0.1.}
\label{fig1}
\end{figure}
\eject

\begin{figure}[htb]
\vspace*{-2cm}
\includegraphics[height=15cm,width=12cm,angle=90]{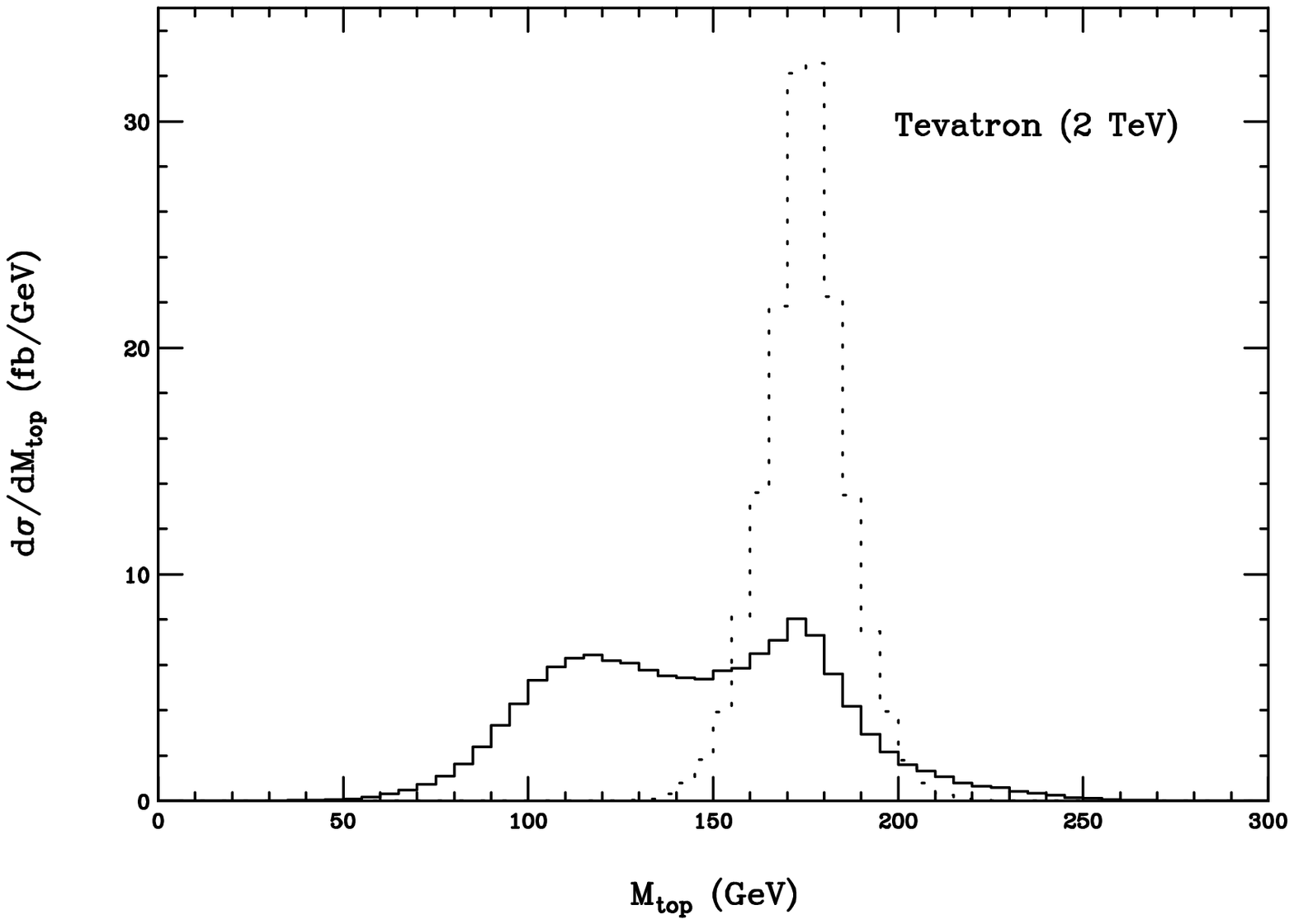}
\caption[]{ The reconstructed top quark mass, $M_{\rm top}(3j)$,
 distribution of  $\ell+4j/b+\ETslash$ at the Tevatron collider. 
The solid curve is for the stop events with stop mass of $170$ GeV.
The dotted curve is for the top quark event scaled down by a factor of 0.1.}
\label{fig2}
\end{figure}
\eject

\begin{figure}[htb]
\vspace*{-2cm}
 \includegraphics[height=15cm,width=12cm, angle=90]{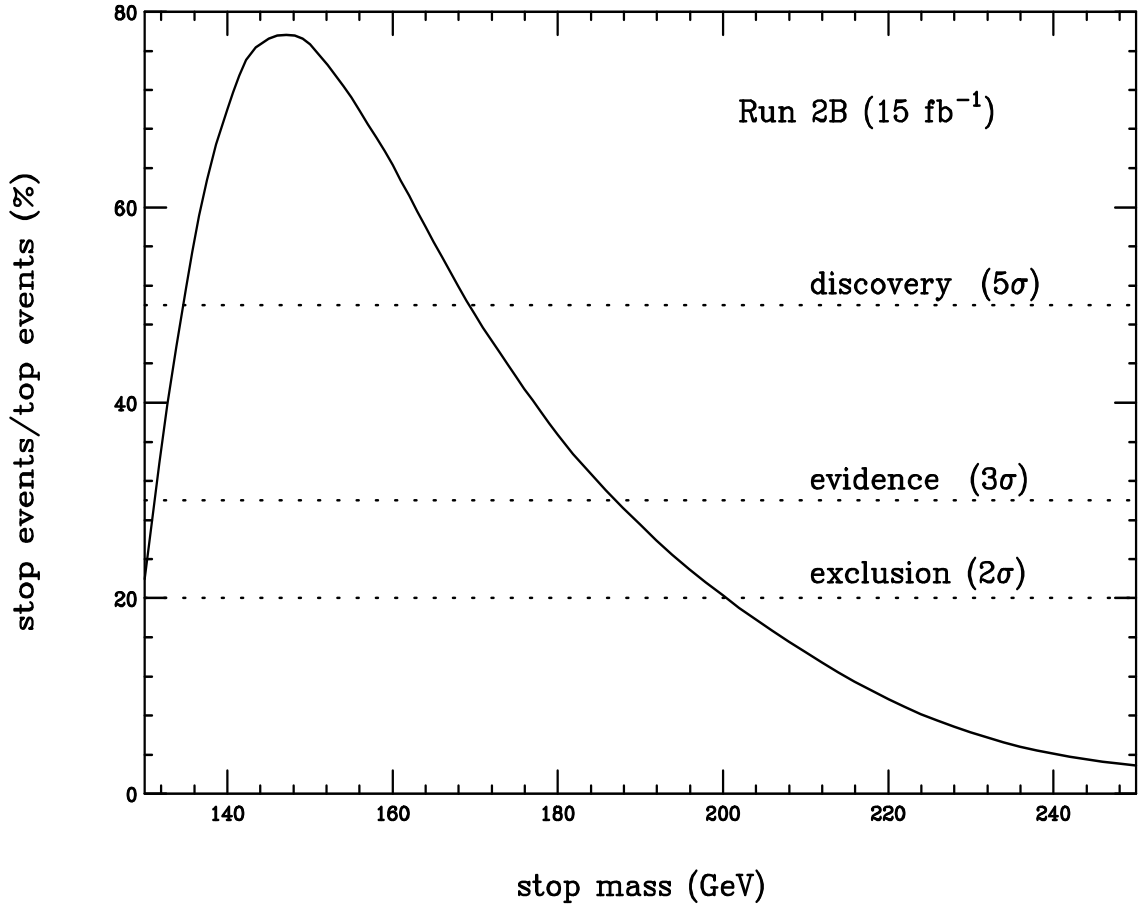}
\caption[]{ The solid curve is the ratio of the stop to top quark event
numbers versus the stop mass under the
basic+$m_T(\ell,p_T^{\rm miss})$+$M_{\rm top}(3j)$
cuts for the upgraded Tevatron (2 TeV).
The three horizontal dotted lines are the discovery, evidence and exclusion 
limits at Run 2B (15 fb$^{-1}$). }
\label{fig3}
\end{figure}
\eject

\begin{figure}[htb]
\vspace*{-2cm}
 \includegraphics[height=15cm,width=12cm, angle=90]{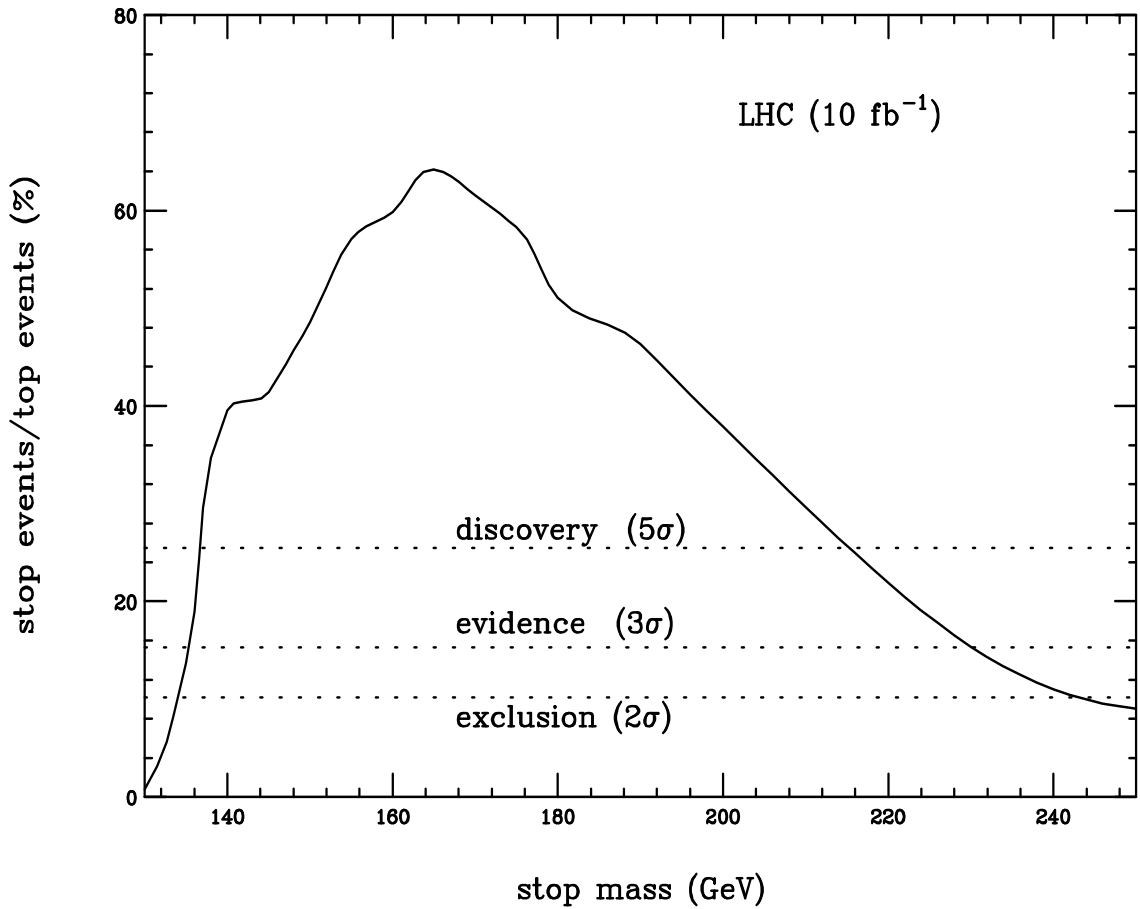}
\caption[]{ The same as Fig. 3, but for the LHC with a luminosity of 
            10 fb$^{-1}$.}
\label{fig4}
\end{figure}
\eject

\begin{thebibliography}{99}
\bibitem{Carena}
	M.~Carena, M.~Quiros and C.E.~Wagner,
	\PLB380, 81 (1996); \NPB503, 387 (1997);
	\NPB524,  3 (1998);
	D.~Delepine, J.M.~Gerard, R.~Gonzalez Felipe and J.~Weyers,
	\PLB386, 183 (1996); J.~McDonald, \PLB413, 30 (1997);
	J.M.~Cline and G.D.~Moore, \PRL181, 3315 (1998).
\bibitem{Dine}M.~Dine, A.~Kagan, and S.~Samuel, \PLB243, 250 (1990);
	S.~Dimopoulos and G.~F. Giudice, \PLB357, 573 (1995); 
	A.~Pomarol and D.~Tommasini, \NPB466, 3 (1996); 
	A.~Cohen, D.~B. Kaplan, and A.~E. Nelson, \PLB388, 599 (1996).  
	See, however, N.~Arkani-Hamed and H.~Murayama, \PRD56, R6733 (1997). 
\bibitem{Hikasa} K. Hikasa and M. Kobayashi, \PRD36, 724(1987).
\bibitem{Hosch}  M. Hosch, R.J. Oakes, K. Whisnant, J. M. Yang, B.-L Young, 
		 X. Zhang, \PRD58, 034002 (1998);
		  S. Mrenna and C.P. Yuan, \PLB367, 188 (1996);
		 Gregory Mahlon, G. L. Kane, \PRD55, 2779 (1997).
\bibitem{Sender} J. Sender, hep-ph/9602354.
\bibitem{CDF} CDF Collaboration, Abstract 652, ICHEP98, Vancouver, July, 1998.
\bibitem{LEP} LEP2 SUSY Working Group, http://www.cern.ch/lepsusy. 
\bibitem{Demina} R. Demina, J. D. Lykken, K. T. Matchev and A. Nomerotski,
		 hep-ph/9910275.
\bibitem{Berger} E. L. Berger and T. M. P. Tait, hep-ph/0002305.
\bibitem{CTEQ5L}  H.~L. Lai, et al., hep-ph/9903282.
\bibitem{zerwas}    W.~Beenakker, M.~Kramer, T.~Plehn, M.~Spira
		    and P.M.~Zerwas, \NPB515, 3 (1998).
\bibitem{report}  For a review, see,  for example, {\it Report of the MSSM
		  Working Group for the Workshop ``GDR-Supersymetrie''},
		  edited by A. Djouadi and S. Rosier-Lees.   
\bibitem{low} For a review, see,  for example, 
	      W. de Boer, R. Ehret, A. V. Gladyshev, D. I. Kazakov,
	      hep-ph/9712376, in the Proc. of ICHEP97, Jerusalem, 1997. 
\bibitem{tev2000} {\it Future ElectroWeak Physics at the Fermilab Tevatron: 
		  Report of the {\rm tev\_2000} Study Group}, edited by D. 
		 Amidei and C. Brock, Fermilab-Pub-96/082;
     A.~P. Heinson, in {\it QCD and High-Energy Hadronic Interactions, 
     Proceedings of the XXXIst Rencontre de Moriond}, Les Arcs, France, 1996, 
     edited by Tran Thanh Van (Edition Frontiere, Gif-sur-Yvette, France, 
    1996),  p.~43 (hep-ex/9605010).
\bibitem{Bonciani} 
   R.~Bonciani, S.~Catani, M.~L. Mangano, and P.~Nason, hep-ph/9801375.
\bibitem{RV} See, for example, 
	     B. Allanach et al. (edited by H. Dreiner), hep-ph/9906224; 
	    S. Abel et al., SUGRA Working Group Collaboration, hep-ph/0003154; 
	     M. Chemtob and G. Moreau,\PRD61, 116004 (2000);
	     P. Chiappetta et al., \PRD61, 115008 (2000);
	     K. Hikasa, J. M. Yang and B.-L. Young, \PRD60, 114041 (1999);
	     R. J. Oakes et al., \PRD57, 534 (1998); 
	     A. Datta, J. M. Yang, B.-L. Young and X. Zhang, 
		   \PRD56, 3107 (1997);
	   D.~K. Ghosh, S.~Raychaudhuri and K.~Sridhar, \PLB396, 177 (1997);
	     E. L. Berger, B. W. Harris and Z. Sullivan, \PRL83,4472 (1999);
	     A. Datta, B. Mukhopadhyaya, hep-ph/0003174.
\end{thebibliography}
\end{document}